# Two-step breakdown of a local $\nu = 1$ quantum Hall state

Masayuki Hashisaka[1-3,*], Koji Muraki[2], Toshimasa Fujisawa[1]

[1]*Department of Physics, Tokyo Institute of Technology, 2-12-1-H81 Ookayama, Meguro, Tokyo 152-8551, Japan*
[2]*NTT Basic Research Laboratories, NTT Corporation, 3-1 Morinosato-Wakamiya, Atsugi, Kanagawa 243-0198, Japan*
[3]*JST, PRESTO, 4-1-8 Honcho, Kawaguchi, Saitama 332-0012, Japan*

We report quantum Hall effect breakdown of a local filling factor $\nu_{local} = 1$ state formed in a bulk $\nu_{bulk} = 2$ system in an AlGaAs/GaAs heterostructure. When a finite source-drain bias is applied across the local system, the breakdown occurs in two steps. At low bias, quantized conductance through the $\nu_{local} = 1$ system breaks down due to inter-edge electron tunneling. At high bias, the incompressibility of the $\nu_{local} = 1$ system breaks down because the spin gap closes. The two steps are resolved by combining measurements of resistively detected nuclear magnetic resonance and shot noise, which allows one to evaluate electron spin polarization in the local system and spin-dependent charge transport through the system, respectively.

DOI:

In integer quantum Hall (QH) systems, electronic current flows along chiral edge channels[1,2]. When a fine gate is used to form a region with local filling factor $\nu_{local} = 1$ so as to traverse a bulk spin-unpolarized $\nu_{bulk} = 2$ system, only spin-up electrons are transmitted through the local region along the $\nu_{local} = 1$ edge channel. The $\nu_{local} = 1$ system, thus operating as an ideal (spin) filter for edge channels at low bias, has been widely used to study charge and spin dynamics in QH systems, such as spin-charge separation[3], providing insights into edge coherence and energy equilibration. When a high bias voltage is applied, however, inter-edge scattering sets in, which drives the $\nu_{local} = 1$ system out of equilibrium so that it exhibits significant nonlinear behavior that bears a resemblance to the QH effect breakdown in macroscopic samples[4,5]. The occurrence of spin-flip scattering as well as spin-conserved scattering in the QH effect breakdown has been manifested by the dynamic nuclear polarization of the host crystal in both integer and fractional regimes[6-9]. Indeed, resistively detected nuclear magnetic resonance (NMR)[10-12] has shown that the electron spin polarization decreases under a high bias in both mesoscopic and macroscopic systems[13]. On the other hand, inter-edge scattering generates shot noise in mesoscopic systems, from which the spin polarization of the transmitted current can be deduced under the assumption that the scattering event is stochastic, i.e., with no correlation[13]. The disagreement between the NMR and shot-noise results found in Ref. 13, in turn, suggests that the charge scattering process generating the shot noise does not directly reflect the electron spin population in the local system.

In this paper, we report a two-step breakdown of a $\nu_{local} = 1$ QH state in a narrow constriction of a $\nu_{bulk} = 2$ system [Fig. 1(a)]. With increasing bias, in the first step, the quantized conductance $G_0 = e^2/h$ (*e*: elementary charge, *h*: Plank's constant) of the $\nu_{local} = 1$ system breaks down due to inter-edge electron tunneling. In the second step, the incompressibility of the $\nu_{local} = 1$ system breaks down because the spin gap closes. This is induced by the suppression of the exchange energy at high bias. We show that the deviation between the NMR and shot-noise results reported in the previous paper[13] appears in the second regime. The well-designed combination of the NMR and shot-noise measurements presented in this paper enables us to gain deep insight into the complicated nonlinear response of the $\nu_{local} = 1$ system, thus opening a way to investigate highly nonequilibrium electron dynamics in mesoscopic QH systems.

A schematic of the experiment is shown in Fig. 1(b). The sample was fabricated in a two-dimensional electron system (2DES) in an AlGaAs/GaAs heterostructure with electron density $n_e = 2.5 \times 10^{11}$ cm$^{-2}$ and mobility $\mu = 3.3 \times 10^6$ cm$^2$V$^{-1}$s$^{-1}$. Electron temperature $T_e$ of the sample was reduced to 14 mK in a dilution refrigerator (base temperature: 7 mK), and an external magnetic field $B_{ext} = 5.0$ T was applied perpendicular to the 2DES to form the $\nu_{bulk} = 2$ state. The magnetic-field direction was from the back to front of the sample so that the chirality of the edge states was clockwise. A split-gate voltage $V_g$ was applied to form a narrow constriction (inset), where a $\nu_{local} = 1$ state is formed. We applied a source-drain voltage $V_{in}$ to drive a current $I_{in}$ through the ohmic contact $\Omega_0$. The dc transport characteristics through the constriction were evaluated by measuring backscattered current $I_1$ through contact $\Omega_1$. In addition to the conductance $G = (I_{in} - I_1)/V_{in} = G_0(T_\uparrow + T_\downarrow)$, we measured the differential conductance $g$

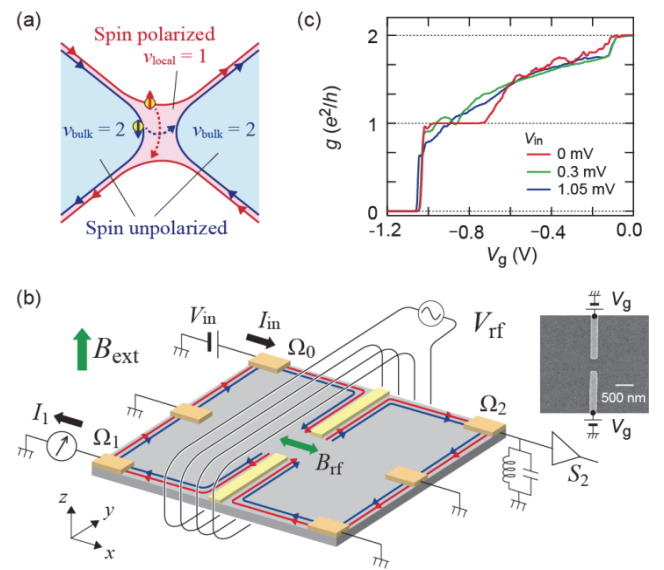

**Figure 1.** (a) Local $\nu_{local} = 1$ QH system formed in a bulk $\nu_{bulk} = 2$ system. (b) Schematic of experimental setup. (Inset) scanning electron micrograph of the device. (c) Pinch-off traces of the split gate measured at several $V_{in}$ values.

= $dI_{in}/dV_{in} - dI_1/dV_{in}$ using a standard lock-in technique with a small ac modulation of $V_{in}$ (15 µV, 37 Hz). Here, $T_{\uparrow(\downarrow)}$ is the transmission probability for spin-up (-down) electrons. The shot noise was evaluated by measuring current fluctuation $S_2 = \langle(\Delta I_2)^2\rangle$ through contact $\Omega_2$ using an LC circuit and a cryogenic amplifier. The resistively detected NMR of $^{75}$As nuclei was measured using a conventional three-step procedure[12-14] by applying a radio-frequency in-plane magnetic field $B_{rf}$ with a four-turn coil. Further details of the measurements are described in **Supplemental material (S.M.)**[15].

The gate voltage $V_g$ dependence of $g$ measured at several $V_{in}$ values is shown in Fig. 1(c). At $V_{in} = 0$ mV, a well-developed $g = e^2/h$ plateau is observed over a wide range of $-1.0$ V $< V_g < -0.7$ V, indicating spin-resolved transport in the lowest Landau level. The $\nu_{local} = 1$ system is formed in this $g = e^2/h$ plateau region. When $V_{in}$ is increased to 0.3 mV, the $g = e^2/h$ plateau disappears, signaling nonlinear transport through the constriction. When $V_{in}$ is further increased to 1.05 mV, the pinch-off trace changes further to show a monotonic decrease in $g$ with decreasing $V_g$.

The nonlinear nature of the breakdown becomes more evident by color plotting $g$ in the $V_{in}$-$V_g$ plane [Fig. 2(a)]. The $g = e^2/h$ plateau, which appears as a yellow area around $V_{in} = 0$ mV extending over $-1.0$ V $< V_g < -0.7$ V, is suddenly terminated at $|V_{in}| = 0.25$ mV. To provide more details, in Fig. 2(b) we plot $g$ traces at several fixed $V_g$ values. The $g = e^2/h$ plateau appears as accumulated traces around $|V_{in}| = 0$ mV. Complex nonlinear behavior is seen at $|V_{in}| > 0.25$ mV. For example, at $V_g = -0.96$ V (red solid curve), $g$ first starts to increase from $e^2/h$ at $|V_{in}| \cong 0.25$ mV and then begins to decrease before it saturates at $g \cong 0.85 e^2/h$ for $|V_{in}| > 0.7$ mV. It should be noted that the nonlinear behavior is seen in the regime where the bias energy $|eV_{in}|$ is considerably larger than the Zeeman energy $E_Z = |g^*\mu_B B|$ ($\cong 0.12$ meV) and much smaller than the exchange-enhanced spin gap $E_Z + E_{ex}$ ($\cong 2$ meV) in bulk 2DES samples[16-18].

The nonlinear behavior in the low-bias regime ($|V_{in}| < 0.45$ mV) can be understood by considering the inter-edge electron tunneling illustrated in Figs. 2(c) and 2(d). The tunneling is manifested in the $V_g$ dependence of $g$. For example, when $V_g$ is slightly increased from $-0.96$ to $-0.90$ V [blue solid line in Fig. 2(b)], $g$ starts to increase from $e^2/h$ at a lower $|V_{in}|$. This indicates the enhanced forward scattering of spin-down electrons, which can be explained by the shorter distance between the $\nu_{bulk} = 2$ edges at a higher $V_g$ [Fig. 2(c)]. When $V_g$ is slightly lowered to $-0.98$ V instead (green solid line), $g$ turns to decrease from $e^2/h$, reflecting the reduced distance between the $\nu_{local} = 1$ edges and the resultant enhanced backscattering of spin-up electrons [Fig. 2(d)]. The inter-edge electron-tunneling picture used here is consistent with the stochastic nature of the shot noise at $|V_{in}| < 0.45$ mV, as discussed later.

To unravel the complex behavior at high bias ($|V_{in}| > 0.45$ mV), we analyze the shot noise and NMR results measured over the entire $|V_{in}|$ range. Figures 3(a), (b), and

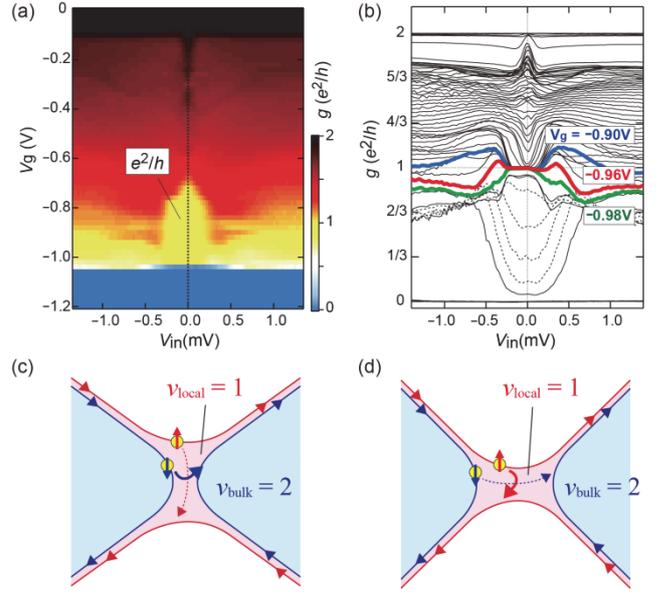

**Figure 2.** (a) Color plot of $g$ as a function of $V_g$ and $V_{in}$. (b) $V_{in}$ dependence of $g$ measured in 20-mV (solid lines: $0 \leq V_g \leq -1.1$ V) and 4-mV (dashed lines: $-1.024 \leq V_g \leq -1.036$ V) steps of $V_g$. Each trace was measured by sweeping $V_{in}$ with a slow speed of 0.067 mV/min. (c)(d) Schematics of the inter-edge tunneling at (c) high and (d) low $V_g$.

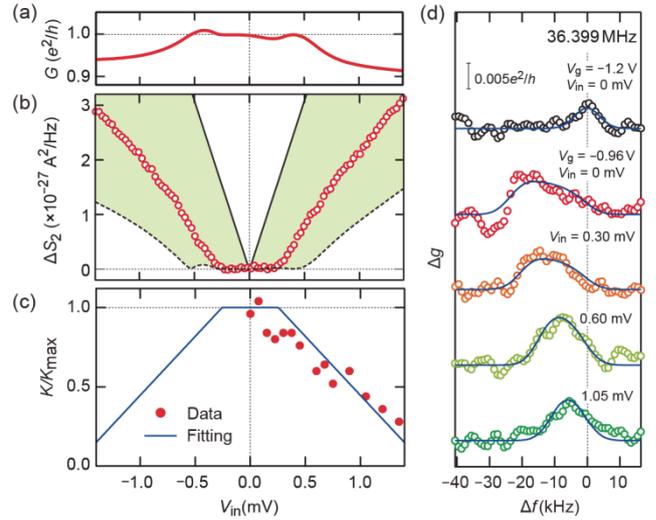

**Figure 3.** (a) $V_{in}$ dependence of $G$ at $V_g = -0.96$ V. (b) $V_{in}$ dependence of $\Delta S_2$. Solid and dashed curves are $S_{shot}$ calculated assuming $F_c = 1$ for fully spin-degenerate and spin-resolved transport, respectively. In general cases, $S_{shot}$ appears in the green area. (c) $V_{in}$ dependence of $K/K_{max}$. (d) Representative NMR spectra. Blue lines are fitted curves.

(c) plot, as a function of $V_{in}$, the results of three different measurements, (a) conductance $G$, (b) excess current noise $\Delta S_2 = S_2(V_{in}) - S_2(0)$, and (c) NMR Knight shift $K$, which were obtained at the same $V_g$ of $-0.96$ V. The nonlinear behavior of $G$, similar to that of $g$, is evident in Fig. 3(a), where $G$ first increases at $|V_{in}| \cong 0.25$ mV and then begins to decrease above $|V_{in}| \cong 0.45$ mV. $\Delta S_2$, which vanishes at low $|V_{in}|$, starts to increase above $|V_{in}| \cong 0.25$ mV [Fig. 3(b)]. $K$ monotonically decreases with increasing $|V_{in}|$ [Fig. 3(c)].

We first compare the observed $V_{in}$ dependence of $\Delta S_2$ with the theoretical shot noise expected for the two limiting cases—spin-degenerate and fully spin-resolved transport—shown as the solid and dashed curves in Fig. 3(b), respectively. Shot-noise power can be expressed as[19]

$$S_{shot} = FS_0 = F \times 2eG_0V_{in}\left[\coth\left(\frac{eV_{in}}{2k_BT_e}\right) - \frac{2k_BT_e}{eV_{in}}\right], \quad (1)$$

where $k_B$ is the Boltzmann constant and the factor $F$ represents the shot-noise reduction due to various mechanisms. To disentangle spin-dependent mechanisms from other ones, here we express $F$ as $F = F_sF_c$, with $F_s = \Sigma_\sigma T_\sigma(1 - T_\sigma)$ representing spin-dependent mechanism ($\sigma = \uparrow$ or $\downarrow$). The factor $F_c$ accounts for other mechanisms, such as the anti-bunching of tunneling electrons. For the moment, we set $F_c = 1$, assuming the electron scattering at the constriction to be stochastic. The $S_{shot}$ vs $V_{in}$ curves for (i) spin-degenerate and (ii) fully spin-resolved transport, shown in Fig. 3(b), were obtained by setting (i) $T_\uparrow = T_\downarrow = G/2G_0$ so that $F_s = 2(G/2G_0)[1 - (G/2G_0)]$ and (ii) $F_s = [G/G_0 - \text{floor}(G/G_0)]\{1 - [G/G_0 - \text{floor}(G/G_0)]\}$, respectively. In general cases, $S_{shot}$ takes intermediate values between the two curves (green area). Indeed, the measured $\Delta S_2$ values fall inside this area.

In the linear-response regime ($|V_{in}| < 0.25$ mV), we observe $\Delta S_2 \cong 0$, in line with spin-resolved transport ($F_s \cong 0$, dashed curve). This unambiguously shows that the transport through the $\nu_{local} = 1$ region is fully spin resolved. As $|V_{in}|$ is increased, $\Delta S_2$ starts to increase at $|V_{in}| = 0.25$ mV. Note that this threshold for finite shot noise is considerably lower than that expected for the spin-resolved transport (0.5 mV, dashed curve). This indicates that the system has entered a different regime where both spin-up and spin-down electrons are involved in the inter-edge scattering. We evaluate the spin polarization $P_T \equiv (T_\uparrow - T_\downarrow)/(T_\uparrow + T_\downarrow)$ of the transmitted current from the shot-noise result. This is possible when the scattering event is stochastic, i.e. $F_c = 1$. We calculate the quantity $\alpha_{shot} = (T_\uparrow' - T_\downarrow')/(T_\uparrow' + T_\downarrow')$, where $T_\uparrow'$ and $T_\downarrow'$ are the quantities obtained by solving the coupled equations $T_\uparrow' + T_\downarrow' = G/G_0$ and $\Sigma_\sigma T_\sigma'(1 - T_\sigma') = \Delta S_2/S_0$. Note that when $F_c = 1$, $T_{\uparrow(\downarrow)}' = T_{\uparrow(\downarrow)}$ so that $\alpha_{shot} = P_T$. In Fig. 4(a), the obtained $\alpha_{shot}$ is plotted as a function of $V_{in}$. The plot reveals the existence of thresholds at $V_{in} \cong 0.25$ mV ($\equiv V_{th1}$) and 0.45 mV ($\equiv V_{th2}$). At $V_{in} = V_{th1}$, $\alpha_{shot}$ starts to decrease from 1 and then, at $V_{in} = V_{th2}$, it stops decreasing linearly and starts saturating toward $\alpha_{shot} \cong 0.9$. We note that $V_{th1}$ and $V_{th2}$ are close to the $V_{in}$ values at which $G$ starts to increase and decrease, respectively [Fig. 3(a)]. Below, we compare $\alpha_{shot}$ with the spin polarization evaluated from NMR measurements.

The Knight shift $K$ of the NMR is proportional to the electron-spin imbalance $\Delta n = n_\uparrow - n_\downarrow$ in the constriction, where $n_\uparrow$ and $n_\downarrow$ are spin-up and spin-down electron densities, respectively[12-14]. Here, we assume $n_\uparrow + n_\downarrow$ to be constant at a given $V_g$ and evaluate the spin polarization $P_n \equiv (n_\uparrow - n_\downarrow)/(n_\uparrow + n_\downarrow)$ at $V_g = -0.96$ V as $P_n = P_{NMR} \equiv K/K_{max}$, where $K_{max}$ is the Knight shift for the fully spin-polarized case. Figure 3(d) shows representative NMR spectra and fitted curves [for the fitting function, see eq. (S1) in **S.M.**[15]]. When the 2DES in the local region is completely depleted at $V_g = -1.2$ V, a peak is observed at 36.399 MHz, which we use as the reference resonance frequency $f_{ref}$ of $^{75}$As nuclei at $B_{ext} = 5.0$ T. The spectra obtained at $V_g = -0.96$ V show finite shifts to lower frequencies. From the fitting at $V_{in} \cong 0$ mV, where the 2DES in the constriction is fully spin polarized, $K_{max} \cong 25$ kHz is obtained. When $V_{in}$ is increased, the NMR spectra shift to a higher frequency toward $f_{ref}$, indicating a decrease in $P_n$ due to the tunneling of spin-down electrons across the constriction. Figure 3(c) summarizes the $V_{in}$ dependence of $P_{NMR}$. We fit the data at $V_{in} > 0.25$ mV by a linear function ($P_{NMR} = -772 |V_{in}| + 1.17$) with the constraint that $P_{NMR} = 1$ in the linear-response regime ($|V_{in}| < 0.25$ mV)[20]. Thus, we obtain $P_{NMR}$ over the entire $V_{in}$ range, as shown by the blue curve.

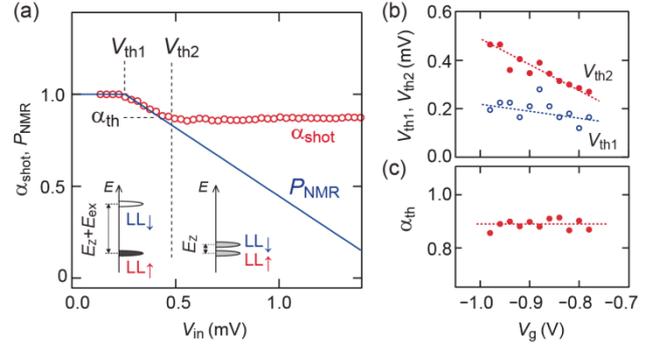

**Figure 4.** (a) $V_{in}$ dependence of $\alpha_{shot}$ and $P_{NMR}$. (Inset) schematics of the lowest Landau levels in the constriction at low- (left) and high-bias (right) regimes. (b) $V_g$ dependence of $V_{th1}$ and $V_{th2}$. (c) $V_g$ dependence of $\alpha_{th}$.

Although $P_T$ and $P_n$ can differ from each other under nonequilibrium conditions, it is reasonable to assume that they have similar $V_{in}$ dependence. We compare the $V_{in}$ dependence of $\alpha_{shot}$ and $P_{NMR}$ (= $P_n$) in Fig. 4(a). At $V_{in} < V_{th2}$, $\alpha_{shot}$ agrees well with $P_{NMR}$, suggesting $\alpha_{shot} = P_T \cong P_n$ in this regime. In contrast, $\alpha_{shot}$ tends to saturate at $\alpha_{shot} \cong 0.9$ for $V_{in} > V_{th2}$, while $P_{NMR}$ monotonically decreases with increasing $V_{in}$. The deviation between $\alpha_{shot}$ and $P_{NMR}$ clearly shows that the electron dynamics at low and high bias are distinct from each other.

Figures 4(b) and 4(c) present the $V_g$ dependence of $V_{th1}$, $V_{th2}$, and $\alpha_{th}$ ($\alpha_{shot}$ value at $V_{in} = V_{th2}$) extracted from the shot-noise results (see **S.M.** for details of the analysis[15]). While $V_{th1}$ and $V_{th2}$ vary depending on $V_g$, $\alpha_{th}$ remains almost constant at $\alpha_{th} \cong 0.9$ independent of $V_g$. This implies that the second step of the breakdown is triggered when the spin polarization drops to $P_n \cong 0.9$ due to spin-down electron tunneling. The decrease in $P_n$ leads to the suppression of the exchange-enhanced spin gap [inset in Fig. 4(a)]. As a result, the incompressible $v_{local} = 1$ state breaks down at $|V_{in}| > V_{th2}$, leading to highly nonequilibrium electron dynamics that is distinct from the inter-edge electron tunneling. The complete disappearance of the $g = e^2/h$ plateau at $V_{in} = 1.05$ mV supports the breakdown picture [Fig. 1(c)]. In addition, the bias-induced exchange-energy suppression can explain why $eV_{th1}$ and $eV_{th2}$ are larger than the Zeeman energy and smaller than the exchange-enhanced spin-gap energy.

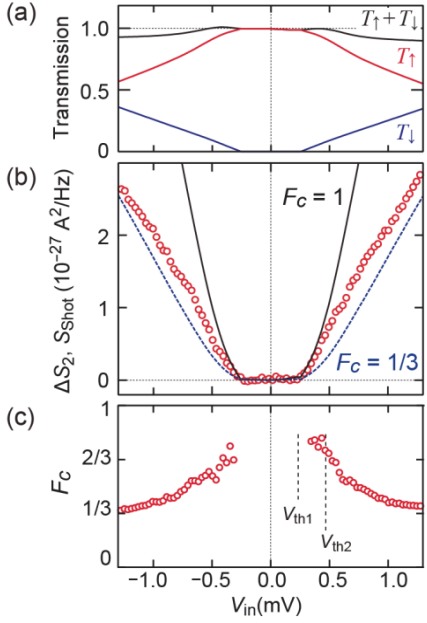

**Figure 5.** (a) $V_{in}$ dependence of $T_\uparrow$, $T_\downarrow$, and $T_\uparrow + T_\downarrow$ estimated assuming $P_{NMR} = (T_\uparrow - T_\downarrow)/(T_\uparrow + T_\downarrow)$. (b) $\Delta S_2$ [same data as in Fig. 3(b)] and $S_{shot}$ curves for $F_c = 1$ and $F_c = 1/3$ calculated using $T_\uparrow$ and $T_\downarrow$ in Fig. 5(a). (c) $V_{in}$ dependence of $F_c$.

We observe that $V_{th2}$ is significantly enhanced with decreasing $V_g$, whereas $V_{th1}$ increases only slightly. The strong $V_g$ dependence of $V_{th2}$ is linked to the $V_g$ dependence of $P_n$. That is, at lower $V_g$, higher $|V_{in}|$ is required to attain a sufficient spin-down tunneling rate [Fig. 3(d)]. This is consistent with the observation that under a finite bias $K$ increases with decreasing $V_g$ (see Fig. S5 in **S.M.**[15]). In contrast, $V_{th1}$ depends on $V_g$ only weakly because it indicates the onset of inter-edge tunneling for either spin-up or spin-down electrons.

In the second breakdown regime, the deviation between $P_{NMR}$ and $\alpha_{shot}$ suggests $F_c < 1$ at $V_{in} > V_{th2}$. In the following, we analyze the shot-noise data at $V_g = -0.96$ V, with the $F_c = 1$ constraint removed and, instead, with the assumption $P_{NMR} = (T_\uparrow - T_\downarrow)/(T_\uparrow + T_\downarrow)$ introduced. With this assumption, $T_\uparrow$ and $T_\downarrow$ can be calculated from the measured $P_{NMR}$ and $G$. The $T_\uparrow$ and $T_\downarrow$ values obtained in this way are plotted in Fig. 5(a) as a function of $V_{in}$, along with $T_\uparrow + T_\downarrow$. We use these $T_\uparrow$ and $T_\downarrow$ traces to calculate $F_s$ as a function of $V_{in}$, which in turn allows one to simulate $S_{shot}$ for arbitrary $F_c$. In Fig. 5(b), we compare the $S_{shot}$ curves simulated for $F_c = 1$ and $F_c = 1/3$ with the experimental $\Delta S_2$ data shown in Fig. 3(b). In this plot, we again observe the two-step behavior of the breakdown: in the low-bias regime ($V_{th1} < |V_{in}| < V_{th2}$), $\Delta S_2$ follows the $F_c = 1$ stochastic shot-noise curve; and at higher bias ($V_{th2} < |V_{in}|$), it deviates from the $F_c = 1$ curve and approaches the $F_c = 1/3$ curve. Figure 5(c) summarizes the $V_{in}$ dependence of $F_c$ obtained from the relation $F_c = \Delta S_2/F_s S_0$. Above $|V_{in}| = V_{th2}$, we find a significant decrease in $F_c$ from $F_c \cong 1$ to $1/3$. Thus, the nonequilibrium transport in the second breakdown regime can be evaluated as a change in $F_c$.

Although electron dynamics in the second nonequilibrium regime is unclear, the shot-noise results provide important insights into the second breakdown process. One possible scenario inferred from the experimental data is the formation of compressible electron liquid. Transport through the compressible liquids of nearly half-filled spin-up and spin-down Landau levels may cause the $F_c \cong 1/3$ shot noise[21-25]. Another possible scenario is the fractional charge tunneling through the local fractional QH system[26-29]. For both scenarios, the incompressibility of the $v_{local} = 1$ state needs to break down. Further experimental and theoretical studies will clarify the electron dynamics in the highly nonequilibrium regime.

In summary, we have investigated the nonlinear behavior of the $v_{local} = 1$ state formed in the $v_{bulk} = 2$ system. A two-step breakdown process was successfully identified through NMR and shot-noise measurements. In the first step, inter-edge electron tunneling breaks down the $v_{local} = 1$ conductance plateau. The second step is caused by the closing of the spin gap due to the suppression of the exchange energy. Shot-noise reduction toward $F_c \cong 1/3$ is observed in the second regime, indicating the breakdown of the incompressible $v_{local} = 1$ state.

We appreciate technical support from T. Endo and fruitful discussions with K. Chida, K. Kobayashi, N. Kumada, and Y. Tokura. This study was supported by Grants-in-Aid for Scientific Research (JP16H06009, JP15H05854, JP26247051), JST PRESTO Grant Number JP17940407, and the Nanotechnology Platform Program (Tokyo Institute of Technology).

# Supplemental Material for
# 'Two-step breakdown of a local $v = 1$ quantum Hall state'

Masayuki Hashisaka[1-3,*], Koji Muraki[2], Toshimasa Fujisawa[1]

[1]Department of Physics, Tokyo Institute of Technology, 2-12-1-H81 Ookayama, Meguro, Tokyo 152-8551, Japan
[2]NTT Basic Research Laboratories, NTT Corporation, 3-1 Morinosato-Wakamiya, Atsugi, Kanagawa 243-0198, Japan
[3]JST, PRESTO, 4-1-8 Honcho, Kawaguchi, Saitama 332-0012, Japan

## 1. Two-terminal resistance of the bulk 2DES.

Figure S1 shows the two-terminal resistance of the 2DES as a function of $B_{ext}$ measured at 30 mK. Signatures of $v_{bulk}$ = 5/3 and 4/3 states are observed at 6.2 and 7.8 T, respectively, as well as the well-developed integer QH plateaus. The breakdown experiment in the main text was performed at $B_{ext}$ = 5.0 T ($v_{bulk} \cong 2.08$), near the center of the $v_{bulk}$ = 2 plateau.

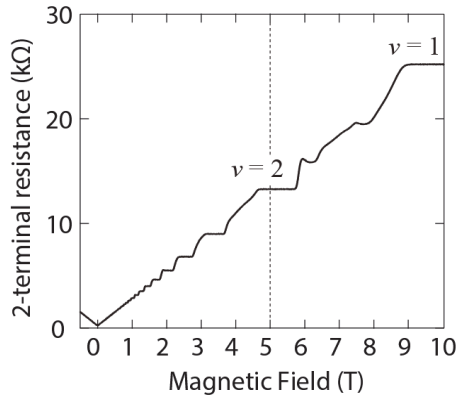

Fig. S1   Magnetic field $B_{ext}$ dependence of the two-terminal resistance of the bulk 2DES.

## 2. DC conductance at several external magnetic fields.

Figure S2(a) shows pinch-off traces of the split gate measured at several $B_{ext}$ values. The $g = e^2/h$ plateau develops above $B_{ext}$ = 2.5 T. Figure S2(b) displays several $g$ traces as a function of $V_{in}$ measured at $B_{ext}$ = 3.3 T. The $g = e^2/h$ plateau is seen only near $V_{in}$ = 0 mV, reflecting the bias-induced breakdown process. On the other hand, the $g = 2e^2/h$ plateau of the $v_{local}$ = 2 state develops over the entire $V_{in}$ range of the measurement, reflecting a large cyclotron gap of about 5 meV.

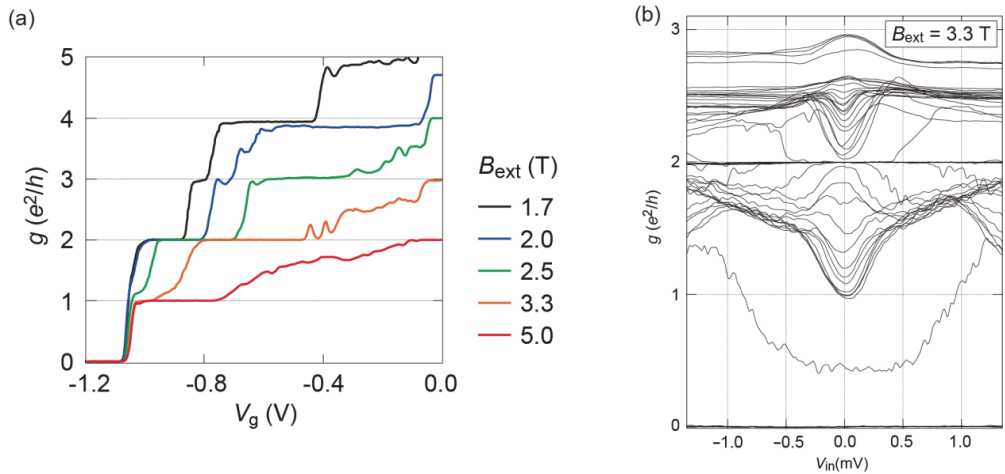

Fig. S2   (a) Pinch-off traces of the split gate at $V_{in}$ = 0 mV measured at several $B_{ext}$. (b) $V_{in}$ dependence of $g$ measured at 3.3 T ($v_{bulk} \cong 3$) in 20-mV steps of $V_g$.

# 3. Experimental techniques.

## A. Resistively-detected NMR.

In our experiment, an NMR spectrum was obtained by measuring the change in differential conductance $\Delta g$ induced by the nuclear spin polarization. This resistively detected NMR is advantageous for investigating local electron and nuclear spin states, because $g$ is only sensitive to charge dynamics in the constriction.

The spin-flip electron scattering induces dynamic nuclear polarization (DNP) near the constriction. This is seen in Figs. S3(a) and (b). Figure S3(a) shows the $V_{in}$ dependence of $g$ measured at $V_g = -0.96$ V with a fast $V_{in}$ sweep rate of 0.53 mV/min. The red and black traces, which are obtained in sweep-up and sweep-down processes, respectively, show significant hysteretic behavior. Figure S3(b) displays the time evolution of $g$ under the temporal binary modulation of $V_{in}$ between 0.42 and $-0.14$ mV. When $V_{in}$ is suddenly changed to $-0.14$ mV after 5 min at 0.42 mV, $g$ gradually increases from $g \cong 0.9 e^2/h$ to $e^2/h$ in a few minutes. This time evolution is explained as follows. Spin-flip electron scattering at $V_{in} = 0.42$ mV induces DNP. When $V_{in}$ is suddenly set to $-0.14$ mV, $g$ is suppressed below $e^2/h$ because of the hyperfine field. Then, $g$ gradually comes back to $e^2/h$ because of the nuclear spin diffusion. The hysteresis and the slow time evolution of $g$ are typical indications of DNP[6-14].

The measurement of the resistively detected NMR was performed using the following three-step procedure[12-14]. First, we fixed the voltages at $V_g = -0.96$ V and $V_{in} = 0.42$ mV and waited for 3 min to prepare DNP. Second, $V_g$ and $V_{in}$ were set at the situation of interest, and the rf voltage $V_{rf}$ ($-26$ dBm) with target frequency $f_{target}$ was applied to the four-turn coil for 5 s to induce in-plane magnetic field $B_{rf}$. Finally, we measured $g$ at $V_g = -0.96$ V and $V_{in} = -0.14$ mV for 50 s and averaged the time-domain data to improve the measurement accuracy. By repeating this procedure with varying $f_{target}$, NMR spectra were obtained as $\Delta f = f_{target} - f_{ref}$ dependence of $\Delta g$ [Fig. 3(d)].

When the 2DES is completely depleted, the NMR peak has a Gaussian profile $\Delta g(\Delta f) \propto \exp(-\Delta f^2/2\gamma^2)$, where $\gamma = 3.79$ kHz is the reference line width, as shown in the uppermost trace in Fig. 3(d). On the other hand, the peak shifts to a lower frequency when the 2DES in the constriction is spin-polarized. The Knight shift is estimated by fitting the measured spectra using the following functions:[14]

$$\Delta g = A \int dy\, dz\, \exp\left\{\frac{-[\Delta f - \Delta f_K(y,z)]^2}{2\gamma^2}\right\} \rho(y,z), \quad (S1)$$

$$\Delta f_K(y,z) = -K \cos^2\left(\frac{\pi y}{\omega_y}\right) \cos^2\left(\frac{\pi z}{\omega_z}\right), \quad (S2)$$

$$\rho(y,z) \propto \cos^2\left(\frac{\pi y}{\omega_y}\right) \cos^2\left(\frac{\pi z}{\omega_z}\right), \quad (S3)$$

where $A$ is the fitting parameter for the peak amplitude, $\Delta f_K(y, z)$ is the position-dependent Knight shift, $\rho(y, z)$ is the electron density in the constriction, and $K = -\alpha_{As}(n_\uparrow - n_\downarrow)\rho(0, 0)$ is the Knight shift at the center of the constriction. Here, $\alpha_{As}$ is the hyperfine coupling coefficient. Note that we define point $(y, z) = (0, 0)$ at the center of the constriction. Under the assumption of confinement widths of $w_y = 65$ nm and $w_z = 18$ nm, the obtained spectra are well fitted by eq. (S1), as shown

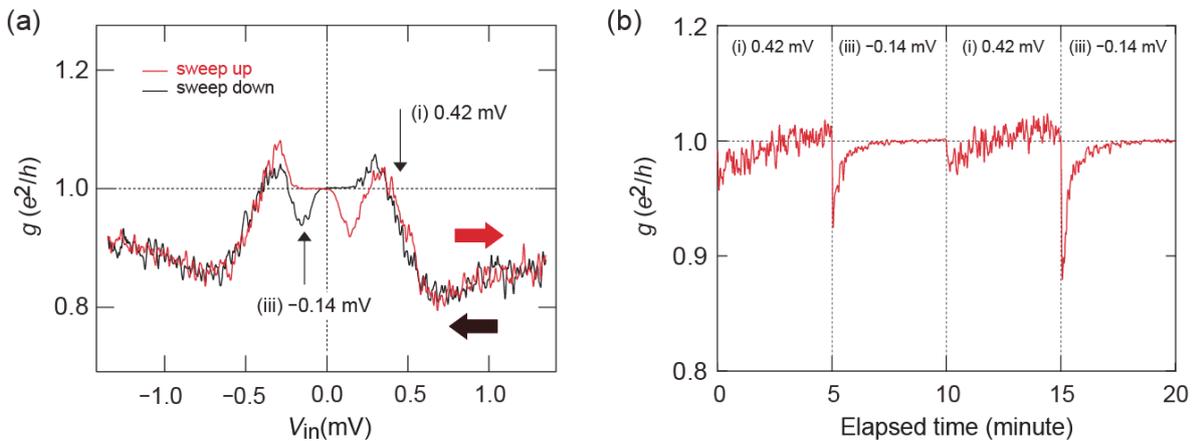

Fig. S3 (a) $V_{in}$ dependence of $g$ measured at $V_g = -0.96$ V with a fast sweep rate of 0.53 mV/min in sweep-up (red) and sweep-down (black) processes. Hysteresis induced by the DNP is the most significant at $|V_{in}| \cong 0.14$ mV. (b) Time evolution of $g$ under the temporal binary modulation of $V_{in}$ between 0.42 and $-0.14$ mV.

in Fig. 3(d). The Knight shift $K_{max} \cong 25$ kHz measured for the fully spin-polarized 2DES is close to the value reported in the previous work[13].

**B. Shot-noise measurement.**

Current noise $S_2$ was measured by a conventional technique using an LC circuit and a homemade cryogenic amplifier[13,29]. $S_2$ was converted to a voltage fluctuation at the 2.4-MHz resonance frequency of the LC circuit and then amplified by the cryogenic amplifier set at 4 K. The output signal was again amplified by a commercial amplifier at room temperature and measured with an analog-to-digital converter.

## 4. Estimations of $V_{th1}$ and $V_{th2}$.

Figure S4(a) displays $\alpha_{shot}$ measured at several $V_g$ values. We estimated $V_{th1}$ and $V_{th2}$ from the inflection points of $\alpha_{shot}$. Figure S4(b) shows the second derivatives of the $\alpha_{shot}$ curves. The inflection points appear as a dip ($V_{th1}$) and a peak ($V_{th2}$) in the data.

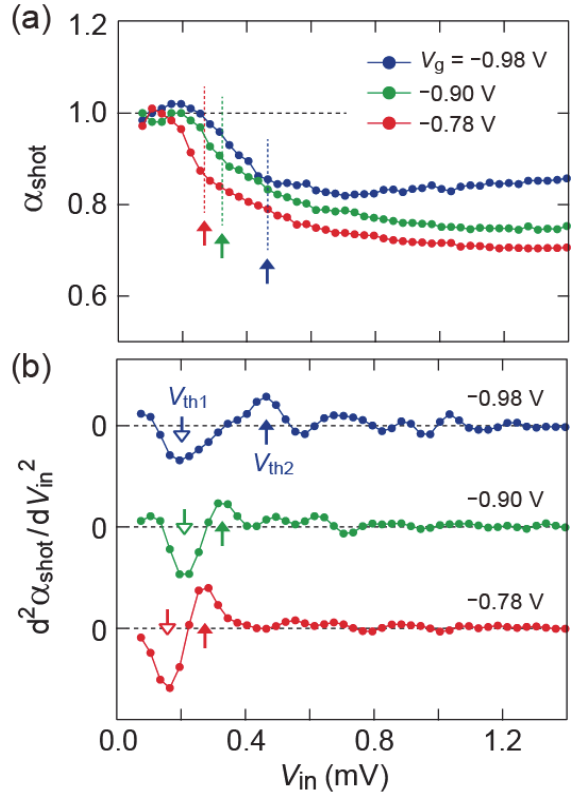

Fig. S4 (a) $V_{in}$ dependence of $\alpha_{shot}$ measured at several $V_g$. (b) Second derivatives of $\alpha_{shot}$ with respect to $V_{in}$. $V_{th1}$ and $V_{th2}$ are obtained as the dip and peak positions, respectively.

## 5. Gate-voltage dependence of the Knight shift and shot noise.

Figures S5(a) and S5(b) show the $V_g$ dependence of $K$ measured at $V_{in} = 0$ and 1.05 mV, respectively. In the second breakdown regime [$-0.8 < V_g < -1.0$ V in Fig. S5(b)], $K$ monotonically increases with decreasing $V_g$ toward $K_{max} = 25$ kHz.

The excess noise $\Delta S_2$ at several $V_g$ values is plotted in Fig. S6. We compare $\Delta S_2$ with $S_{shot}$ calculated assuming $F_c = 1$ for spin-degenerate (black solid curve) and spin-resolved (black dashed) transport. In addition, we plot $S_{shot}$ calculated assuming $F_c = 1/3$ for spin-degenerate transport (blue curve). At $V_g = -0.92$ V, in the low-bias regime ($0.25 \lesssim |V_{in}| \lesssim 0.4$ mV) $\Delta S_2$ develops following the spin-polarized $F_c = 1$ curve (black dashed), while at high bias it starts to deviate from the curve and approaches the spin-degenerate $F_c = 1/3$ curve (blue). This supports the two-step breakdown scenario, the same as the results at $V_g = -0.96$ V do (main text). Similar shot-noise reduction was observed over the wide range of $V_g$, for example, at $V_g = -1.024, -1.028,$ and $-1.036$ V, where $g < e^2/h$ at $V_{in} = 0$ mV.

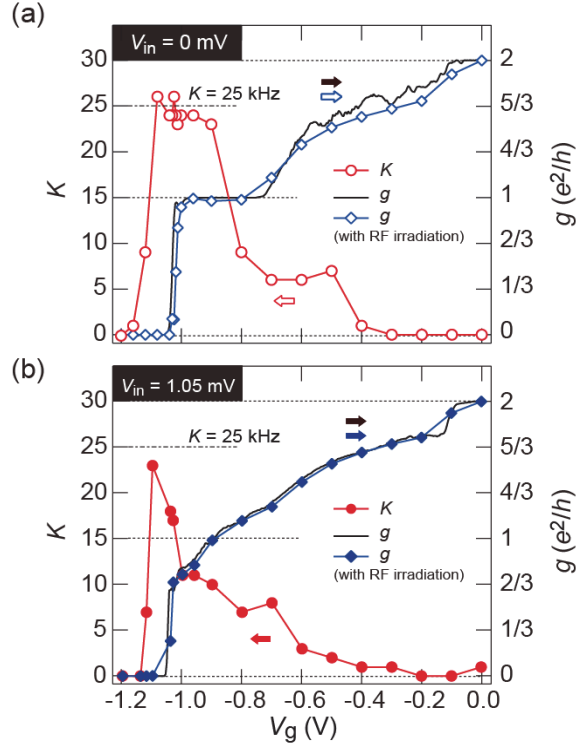

Fig. S5 (a)(b) $V_g$ dependence of $K$ at $V_{in} = 0$ and 1.05 mV, plotted with pinch-off traces of the constriction with and without $B_{rf}$ irradiation.

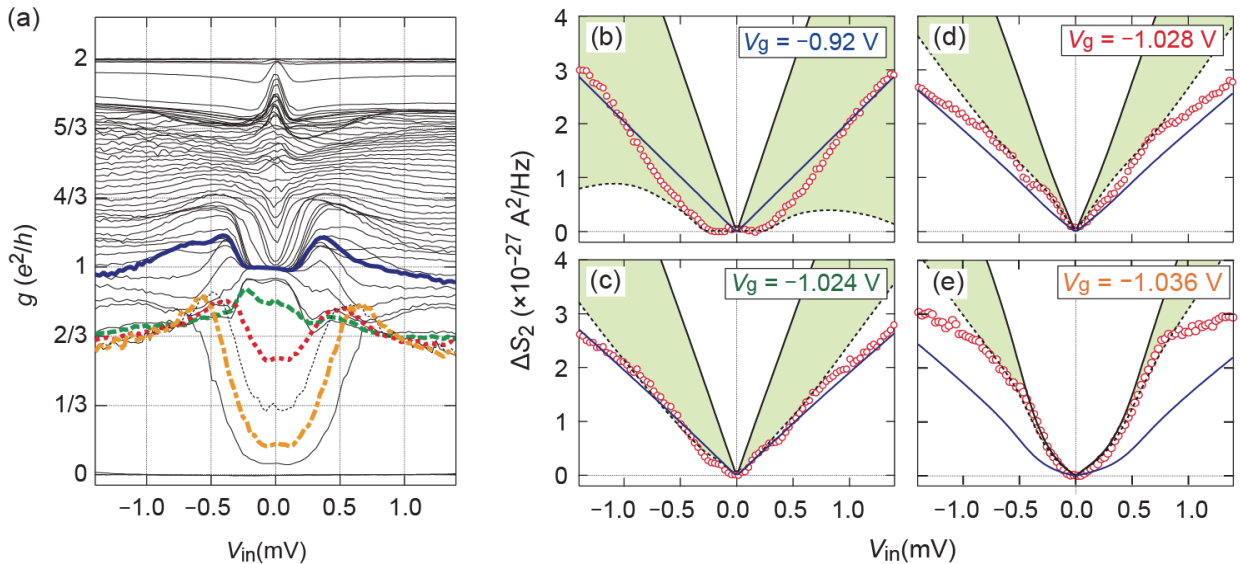

Fig. S6 (a) $V_{in}$ dependence of $g$ measured in 20-mV (solid lines) and 4-mV (dashed lines) steps for $V_g$ [same data as in Fig. 2(b)]. (b)(c)(d)(e) $V_{in}$ dependence of $\Delta S_2$ measured at $V_g = -0.92, -1.024, -1.028,$ and $-1.036$ V. Black solid and dashed lines are $S_{shot}$ calculated for $F_c = 1$, assuming fully spin-degenerate and spin-resolved transport, respectively. Green region indicates possible $S_{shot}$ values with $F_c = 1$. Blue solid line indicates $S_{shot}$ calculated for $F_c = 1/3$, assuming fully spin-degenerate transport.